\documentclass[prd,nofootinbib,onecolumn]{revtex4}
\usepackage{amsmath}
\usepackage{amssymb}
\usepackage{graphics}
\usepackage{epsfig}

\newcommand\nn{\nonumber}
\newcommand\ba{\begin{eqnarray}}
\newcommand\ea{\end{eqnarray}}

\begin{document}

\title{Azimuthal correlation of gluon jets created in proton-antiproton annihilation}

\author{E.~A.~Kuraev}
\email{kuraev@theor.jinr.ru}
\affiliation{Joint Institute for Nuclear Research, Dubna, Russia}

\author{V. V. Bytev}
\email{bytev@theor.jinr.ru}
\affiliation{Joint Institute for Nuclear Research, Dubna, Russia}
%

\author{E. S. Kokoulina}
\email{kokoulin@sunse.jinr.ru}
\affiliation{Joint Institute for Nuclear Research, Dubna, Russia}

\date{\today}

\begin{abstract}
Annihilation process of proton and antiproton to
guark antiquark pair accompanied by emission of two additional
gluon jets with intermediate state of vector meson is considered.
Strong azimuthal correlation is revealed of two gluonic jets,
effectively created in the same plane.
Some applications to cosmic ray events and to LHC experiments are
discussed.

\end{abstract}

\maketitle

\section{Introduction}

Now is widely believed that hard processes with hadron interaction
at high energies can be described in frames of QCD. Contrary to
photons in QED the QCD gluons can interact between themselves. The
experimental check of nature of gluons become an important problem
up to now. The dominant role of branching gluons was utilized  in
construction of the gluon dominance model (GDM) describing the
events with large multiplicity in high energy collisions of leptons,
(anti)protons and ions \cite{kok}. Another indirect indication can
be found in measuring the azimuthal correlations between gluon jets
emitted by color quarks created in some hard hadronic process. It is
the motivation of this paper to show that such kind of correlation
take place in particular in process of annihilation of the high
energy proton and antiproton to quark-anti quark with emission of
two gluons.

Years ago in the paper of one of authors \cite{1978} a problem of
inelastic quark form factor was considered. The consideration was
performed in the so called double-logarithmical approximation (DL)
when the leading terms of order $(\alpha_s/\pi)^2(\ln(Q^2/m^2))^4$
was taken into account (here $Q^2=-q^2$ is the square of 4-momentum
of virtual vector particle, $m$ is the effective mass of the gluon
jet and $\alpha_s$ is QCD coupling constant). The one-loop radiative
correction to the inelastic form-factor with a single hard gluon
emission was found as well as the contribution from the channel of
two hard gluons creation. It was confirmed, in particular, the S.
Adler statement about cancelation of all DL enhancements when
considering two-loop virtual corrections together with contribution
of inelastic channels. Compared with QED case apart from some
complication connected with the color properties of quarks and
gluons, some additional kinematical regions arising from the
existence of three gluon interaction vertex become to be important.
It is the motivation of this paper to generalize the results
obtained in 1978 year to the annihilation channel. Another reason is
to search the possible relation with some azimuthal correlation of
particles observed at LHC \cite{2010}.

Some additional speculation can be done, a prediction about the
character of energy distribution of cosmic rays of high energy
\cite{cosmic}. Really, taking into account the strong azimuthal
correlation of jets created in high energy proton with nuclei in
atmosphere (all the jets are in the same plane) a specific
distribution of the spots of sets of the secondary particles of high
energy can be measured: all these spots must be located along some
lines-intersection line of production plane and the surface of the
Earth.

Process of annihilation of proton and anti-proton to a vector meson with the
subsequent creation of quark-anti-quark pair and two hard gluon jets
\ba
p(p_+)+\bar{p}(p_-) \to v(q)\to q(q_-)+\bar{q}(q_+)+g(k_1)+g(k_2)
\ea
in Born approximated is described by a set of Feynman diagrams of two kinds.
One of them corresponds to
emission of both gluons by quark and anti-quark and another contain the specific for
QCD three gluon vertex, when two gluons arise from a single gluon, emitted by quark or
anti-quark.
We will use Sudakov parametrization for 4-momenta of the problem, introducing two light-
like vectors
\ba
k_i=p_1\beta_i+p_2\alpha_i+k_{i\bot},k_{i\bot}p_{1,2}=0; k_{i\bot}^2=-\vec{k}_i^2<0; \nn \\
p_1^2=p_2^2=0; s=(q_++q_-)^2, p_1=q_--\frac{1-\beta}{1+\beta}q_+;p_2=q_+-\frac{1-\beta}{1+\beta}q_-; \nn \\
q_+^2=q_-^2=M^2=\frac{s}{4}(1-\beta^2); \nn \\
k_i^2=s\alpha_i\beta_i-\vec{k}^2=m^2.
\ea
Here $M$ is quark mass and $m$ is the mass of vector particle - gluon (or the invariant mass of the
relevant gluon jet).

In terms of Sudakov variables the phase volume of two real soft gluons can be written in form
\ba
\frac{d^3 k_1}{2\omega_1}\frac{d^3 k_2}{2\omega_2}=d^4 k_1 \delta(k_1^2-m^2)d^4 k_2 \delta(k_2^2-m^2)= \nn \\
\frac{s}{2}d\alpha_1 d\beta_1 d^2\vec{k}_1\delta(s\alpha_1\beta_1-\vec{k}^2-m^2)
\frac{s}{2}d\alpha_2 d\beta_2 d^2\vec{k}_2\delta(s\alpha_2\beta_2-\vec{k}^2-m^2)= \nn \\
(\frac{s}{2})^2\frac{d\phi_1}{2}d\alpha_1 d\beta_1 \theta(\alpha_1\beta_1-\gamma)\frac{d\phi_2}{2}d\alpha_2 d\beta_2 \theta(\alpha_2\beta_2-\gamma),
\ea
with $\gamma=m^2/s<<1$.
For the ratio of matrix elements squared  and summed on  final state quantum numbers was obtained in \cite{1978}:
\ba
\frac{\sum|M^{p\bar{p}\to q\bar{q}gg}|^2}{\sum|M^{p\bar{p}\to q\bar{q}}|^2}=(\frac{\alpha_s}{\pi})^2\times Z; \nn \\
Z=\{(\frac{1}{\alpha_2\beta_1}+\frac{1}{\alpha_1\beta_2})\frac{1}{(\beta_1+\beta_2)(\alpha_1+\alpha_2)}
(\frac{N^2-1}{2N})^2
\nn \\
+(-\frac{N^2-1}{4N^2})\{(\frac{1}{\alpha_2\beta_1}+\frac{1}{\alpha_1\beta_2})
\frac{1}{(\beta_1+\beta_2)(\alpha_1+\alpha_2)}+
\frac{1}{\alpha_1\alpha_2\beta_1\beta_2}\}
\nn\\
+\frac{(N^2-1)s}{4((k_1+k_2)^2-m^2)}
\{\frac{1}{\alpha_2\beta_1}+\frac{1}{\alpha_1\beta_2}+\frac{1}{(\beta_1+\beta_2)(\alpha_1+\alpha_2)}
\{-4+
\frac{\alpha_1}{\alpha_2}+\frac{\alpha_2}{\alpha_1}+\frac{\beta_1}{\beta_2}+\frac{\beta_2}{\beta_1}
\}\}\}
\nn\\
=A+\frac{s}{(k_1+k_2)^2-m^2}B.
 \ea
$N=3$ is the rank of color group $SU(N)$. Main contribution arises
from two kinematical configurations in 4-dimensional manyfold $R$:
$\gamma<\alpha_i,\beta_i<1, \alpha_i\beta_i>\gamma$ with the
additional conditions  in the region $a$)
$$\alpha_i>>\alpha_j, \quad \beta_i>>\beta_j, \quad i,j=1,2$$
and region $b)$
$$1>>|\alpha_1\beta_2-\alpha_2\beta_1|>>\gamma.$$

Contribution of first kind kinematics, region $a)$, is \ba
\frac{d\sigma^{p\bar{p}\to q\bar{q}gg}}{d\sigma_0}|_{a)}=
\frac{1}{2!}(\frac{\alpha_s}{2\pi})^2\rho^4c_F^2, \quad
\rho=\ln(s/m^2), \quad c_F=\frac{N^2-1}{2N} \ea where $\alpha_s$ is
gluon coupling constant and $d\sigma_0$ is the differential cross
section of annihilation of proton and anti-proton to the pair of
quark and anti-quark (see Appendix A).

But it is not a total answer. The double-logarithmic type
contribution arises as well from the region $b)$. Really the
denominator of the virtual gluon propagator \ba
(k_1+k_2)^2-m^2=2k_1k_2+m^2=s(\alpha_1\beta_2+\alpha_2\beta_1)+m^2-2|\vec{k}_1||\vec{k}_2|
\cos\phi=s(a-b\cos\phi),
\nn \\
\vec{k}_i^2=s\alpha_i\beta_i-m^2. \ea Being averaged on azimuthal
angle $\phi$  it has a form \ba
\int\limits_{-\pi}^{\pi}\frac{d\phi}{2\pi}\frac{1}{a-b\cos\phi}=\frac{1}{\sqrt{a^2-b^2}}.
\ea Using the on mass shell conditions for gluons we have \ba
\frac{1}{\sqrt{a^2-b^2}}=\frac{1}{\sqrt{(\alpha_1\beta_2-\alpha_2\beta_1)^2+4\gamma[\alpha_1\beta_2+\alpha_1\beta_1+\alpha_2\beta_2]}}.
\ea We see that in region $b)$ it is presented a quite different
region of DL contributions to the cross section.

Performing the integration it was obtained ( \cite{1978} expression
(17)): \ba
\frac{d\sigma^{(2)}}{d\sigma_0}=\frac{1}{2!}c_F(\frac{\alpha_s}{2\pi}\rho^2)^2[c_F+\frac{1}{6}c_V].
\ea Color structure containing $c_V=N$ violates Poisson distribution
in emission of two colored vector particles \ba
\frac{d\sigma^{(n)}}{d\sigma_0}=\frac{1}{n!}a^n e^{-a}, \quad
a=\frac{\alpha_s c_F \rho^2}{2\pi}. \ea Here factor with $exp(-a)$
takes into account the contributions from virtual vector meson
emission.

We suggest to measure the azimuthal correlation in form
\ba
A(\phi)=\frac{d\sigma^{(2)(\phi)}}{\int\limits_{-\pi}^\pi d\sigma^{(2)}(\phi) d\phi}=\nn \\
\frac{c_F^2+8c_Fc_V
\pi\frac{Z(\rho,\phi)}{\rho^4}}{2\pi(c_F^2+\frac{1}{6}c_Fc_V )}=
\frac{4+72\pi \frac{Z(\rho,\phi)}{\rho^4}}{11 \pi}. \ea Function
$Z(\rho,\phi)$ satisfies the normalization condition
$\int\limits_{-\pi}^\pi A(\phi) d \phi=1$

To obtain $Z(\rho,\phi)$ we simplify the expression $a-b\cos\phi$ in
such a way: \ba
\frac{1}{a-b\cos\phi}\approx\frac{2a}{a^2-b^2+a^2\phi^2}, \quad
a=2\alpha_1\beta_2. \ea Performing the $\alpha_2$ integration we
obtain \ba
\int\frac{d\alpha_2}{a-b\cos\phi}=\frac{4\pi\alpha_1\beta_2}{\beta_1^2}\frac{1}{\sqrt{R}}.
\ea The quantity $R$ have a different form depending on integration
regions. We obtain (see Figs. 2--4): \ba
Z(\rho,\phi)=\int\limits_{m/\sqrt{s}}^1\frac{d\beta_1}{\beta_1}\int\limits_{m^2/(s\beta_1)}^{\beta_1}
\frac{d\alpha_1}{\alpha_1}\times
\{\int\limits_{\sqrt{m^2\beta_1/(s\alpha_1)}}^{\beta_1}\frac{d\beta_2}{\sqrt{R_1}}+
 \int\limits_{\beta_1}^1\frac{d\beta_2}{\sqrt{R_2}}\}
+\int\limits_{m/\sqrt{s}}^1\frac{d\alpha_1}{\alpha_1}\int\limits_{m^2/(s\alpha_1)}^{\alpha_1}
\frac{d\beta_1}{\beta_1}\times
\{\int\limits_{\sqrt{m^2\beta_1/(s\alpha_1)}}^{\beta_1}\frac{d\beta_2}{\sqrt{R_1}}+ \nn \\
\int\limits_{\beta_1}^{\beta_1/\alpha_1}\frac{d\beta_2}{\sqrt{R_2}}\}.
\ea
with
\ba
R_1=\frac{4m^2}{s}\frac{\alpha_1}{\beta_1}+4(\frac{\alpha_1\beta_2\phi}{\beta_1})^2; \nn \\
R_2=\frac{4m^2}{s}\frac{\alpha_1\beta_2^2}{\beta_1^3}+4(\frac{\alpha_1\beta_2\phi}{\beta_1})^2.
\ea
We use the integrals
\ba
\int\limits_{\beta_1}^1\frac{d\beta_2}{\sqrt{R_2}}=\frac{\beta_1}{2\alpha_1}\frac{\ln(1/\beta_1)}{\sqrt{\phi^2+\frac{m^2}{s\alpha_1\beta_1}}} \nn \\
\int\limits_{\beta_1}^{\beta_1/\alpha_1}\frac{d\beta_2}{\sqrt{R_2}}=\frac{\beta_1}{2\alpha_1}\frac{\ln(1/\alpha_1)}{\sqrt{\phi^2+\frac{m^2}{s\alpha_1\beta_1}}}
\ea
and
\ba
\int\limits_{\sqrt{\gamma\beta_1/\alpha_1}}^{\beta_1}\frac{d\beta_2}{\sqrt{R_1}}=\frac{\beta_1}{2\alpha_1|\phi|}\times \nn \\
\{\ln(t|\phi|+\sqrt{1+t^2\phi^2})-\ln(|\phi|+\sqrt{1+\phi^2})\}, t=\sqrt{\alpha_1\beta_1/\gamma}.
\ea
To express the result in terms of one-fold integrals we use
\ba
\int\limits_{\sqrt{\gamma}}^1\frac{d\beta_1}{\beta_1}\int\limits_{\gamma/\beta_1}^{\beta_1}\frac{d\alpha_1}{\alpha_1}F(t)=
\frac{1}{2}\int\limits_0^\rho dy (\rho-y)F(\exp(y/2)); \nn \\
\int\limits_{\sqrt{\gamma}}^1\frac{d\beta_1}{\beta_1}\ln(1/\beta_1)\int\limits_{\gamma/\beta_1}^{\beta_1}\frac{d\alpha_1}{\alpha_1}F(t)=
\frac{1}{8}\int_0^\rho dy (\rho-y)^2F(\exp(y/2)); \nn \\
\int\limits_{\sqrt{\gamma}}^1\frac{d\beta_1}{\beta_1}\int\limits_{\gamma/\beta_1}^{\beta_1}\frac{d\alpha_1}{\alpha_1}=
\frac{1}{2}\int_0^\rho dy (\rho-y)=\frac{1}{4}\rho^2. \ea The result
is \ba
Z(\rho,\phi)=\frac{1}{4}\int\limits_0^\rho (\rho-y)^2\frac{1}{\sqrt{\phi^2+e^{-y}}}d y+ \nn \\
\frac{1}{|\phi|}\int\limits_0^\rho(\rho-y)\ln(|\phi|e^{y/2}+\sqrt{1+\phi^2e^y})d
y+O(\rho^2). \ea Function $Z(\rho,\phi)$ has a delta-function type
behavior, concentrated in $|\phi|\sim \sqrt{s/m^2}$ with
$Z(0)\approx 8\sqrt{s/m^2}$ and besides
$\int\limits_{-\epsilon}^\epsilon Z(\phi) d\phi=(1/24)\rho^4$. Here
$\sqrt{m^2/s} << \epsilon <<1.$

This function is presented in Figure 1 for $\rho=10$.

\section{Discussion}

Let remind the three dimensional picture in center of mass of
initial particles. Hard quark and anti-quark of final state move
back to back at large angles and emit two gluons. One of them moves
close to quark direction another - close to anti-quark ones.
Differential cross section on the azimuthal angle between the planes
containing quark and the relevant gluon and the plane containing
anti-quark and another gluon will have a isotropic part and one with
sharp dependence  distributed close to the value $|\phi|\approx
O(\sqrt\frac{m^2}{s})$.

Energies of gluons are small compared with energies of quark-and
anti-quark. This statement follows from the fact that the
main,($\sim \rho^4$) contribution follows from the 4-vectors of
gluons polarization in kinematics of isotropic contributions to the
cross section are situated in the plane of quark and anti-quark
4-momenta
$e_i=-\sqrt{\frac{\alpha_i}{s\beta_i}}p_2+\sqrt{\frac{\beta_i}{s\alpha_i}}p_1$,
whereas the ones, responsible for azimuthal correlation are
essentially situated in the plane transversal to quark momenta.
\begin{figure}
\begin{center}
\includegraphics[scale=.8]{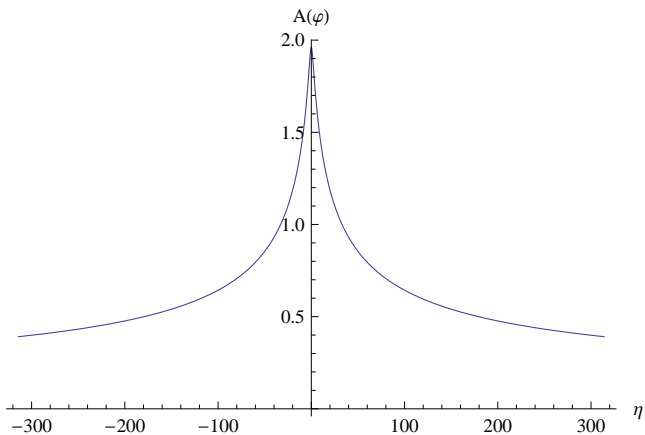}
\end{center}
\caption{the azimuthal distribution function
$A(\phi)$ for $m^2/s=10^{-4}$, $\phi=\eta\sqrt{m^2/s}$ is presented.}
\label{fig1}
\end{figure}
For the case of production of additional soft quark-anti-quark
instead of two gluon the DL enhancement factor do not appears.

The DL enhancement phenomena will take place for gluons emitted by any pair of colored fermions
which have large invariant mass $\sqrt{s}>>m$. However the explicit form of $\phi$-independent
and $\phi$-dependent parts of matrix element square (see (4)) will depend on mechanism of quark
creation. In particular one can consider the peripheral mechanism of
quark-anti-quark creation in collisions of hadrons
by two reggeized gluons.

\begin{figure}
\begin{center}
\includegraphics[scale=.8]{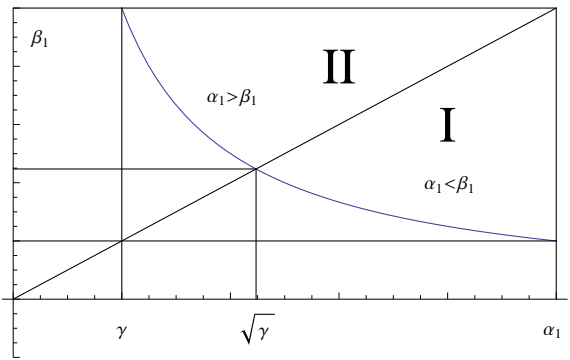}
\end{center}
\caption{Integration domain in $\alpha_1$, $\beta_1$ plane.}
\label{fig12}
\end{figure}

\begin{figure}
\begin{center}
\includegraphics[scale=.8]{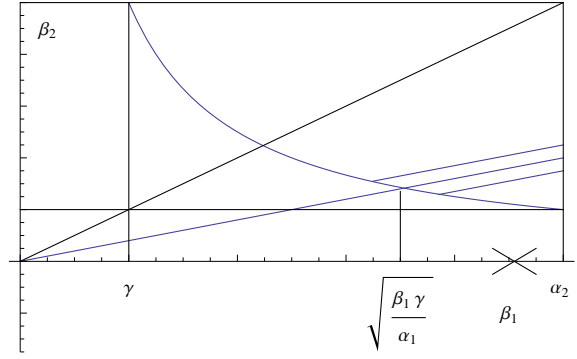}
\end{center}
\caption{Integration domain  for I region  (see Fig.2 )in $\alpha_2$, $\beta_2$ plane.}
\label{fig3}
\end{figure}

\begin{figure}
\begin{center}
\includegraphics[scale=.8]{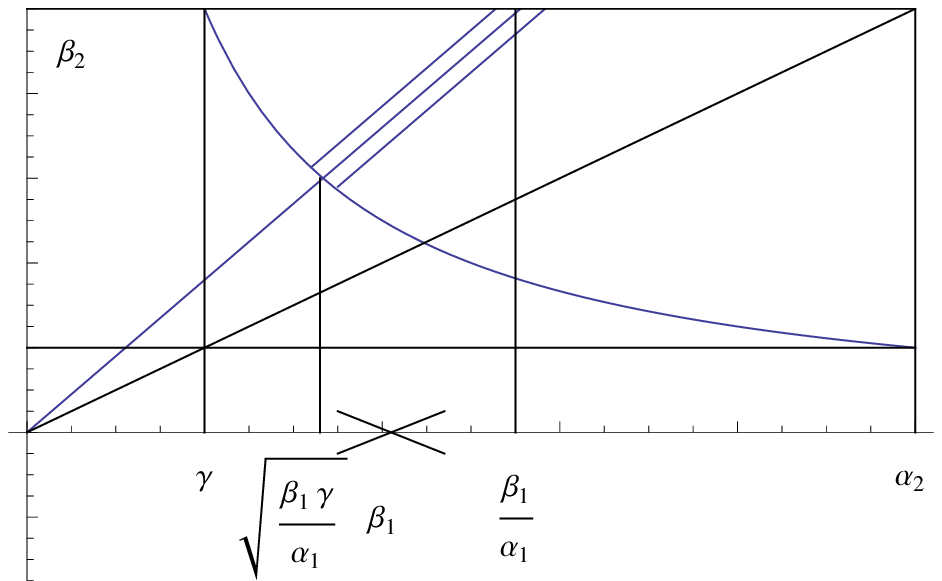}
\end{center}
\caption{Integration domain in $\alpha_2$, $\beta_2$ plane for II region (see Fig.2).}
\label{fig4}
\end{figure}

From the comparisons with experimental topological cross sections of
$pp$ interactions and  $p\bar p$ annihilation at the energy region
close to U-70 accelerator (IHEP, Protvino) the estimations of the
MGD-parameters were obtained \cite{kok}. The mean number of hadrons
formed from one gluon source at hadronization stage is added from
the charged and neutral components (pions predominate): $\overline
n_{tot}$=1.63+1.01=2.64. The invariant mass of such gluon sources
may be determined as $m$= 2.64 $\times $ 0.139= 0.37 (GeV). At
accelerator U-70 energy  $\sqrt s$=11.6 GeV. The parameter $\rho $
will be equal to, $\rho =\ln \frac{s}{m^2}$=6.9. At that $\rho $ the
noticeable peak may appear in the angular dependence too. It looks
also as 1-GeV g-jet at low threshold of ISR-energy, $\sqrt s$=32
GeV. To describe the topological cross section widening at ISR
energies the gluon fission was included at QCD-cascade stage in GDM
\cite{kok}. One can make next assumptions. To reveal the angular
distribution peak appeared from gluon fission at the energy lower
than LHC region it is enough to choose g-jets with smaller invariant
mass m or determine mass of g-jet by means the peak. Also the
estimation of the invariant gluon mass is compared to the mean
transverse momentum of secondary particles at the relevant energies
by the surprising way and this can be clue to ridge phenomenon
understanding.

\section{Appendix A}

Matrix element of process $p(p_+)+\bar{p}(p_-) \to v(q)\to q(q_-)+\bar{q}(q_+)$ have a form
\ba
M_0=\frac{g_{Vpp}g_{Vqq}F_q(q^2)}{q^2-M_V^2}\bar{v}(p_-)\Gamma_\mu u(p_+)\times\bar{u}(q_-)\gamma_\mu v(q_+)
\ea
with phenomenal approach for the quark form factor
and $g_{Vpp}=3g_{Vqq}\approx 9$, $V$ is dominantly $\omega$-meson.

Differential cross section have a form ($c=\cos\theta$, $\theta$ is the center of mass angle
between the directions of motion of initial proton and the positively charged quark
with momentum $q_+$)
\ba
\frac{d\sigma_0}{d c}=\frac{1}{2}\sigma_0; \nn \\
\sigma_0=\frac{N(G_pG_q)^2}{2\pi s\beta(s-M_V^2)^2}\beta_q R_0, \beta_q=\sqrt{1-\frac{4m_q^2}{s}}, \nn \\
R_0=R_1|F_m|^2+\frac{2}{1-\tau}(|F_m|^2-|F_e|^2)R_2 ,
\ea
with $N=3$ is the number of quark colors
\ba
F_m=F_1+F_2; F_e=F_1+\tau F_2, \tau=\frac{s}{4M_p^2}
\ea
and
\ba
R_1=t^2+u^2+4s(M^2+m_q^2)-2(M^2+m_q^2)^2; \nn \\
R_2=tu+s(m_q^2-(M^2+m_q^2)^2.
\ea
Here we use the kinematic invariants
\ba
s=(p_++p_-)^2;t=(p_+-q_+)^2; u+(p_+-q_-)^2; s+t+u=2(M^2+m_q^2),
\ea
$M,m_q$-proton and quark masses.
In paper \cite{F2} was argued that dominant contribution is provided by $V$-omega meson,
and, besides $F_2^\omega=0$.
In paper \cite{Fp} the reasonable model for Dirac formfactor of proton in annihilation channel was
suggested
\ba
F_1(s)=\frac{\Lambda^4}{\Lambda^4+(s-M_p^2)^2}.
\ea

\section{acknowledgements}

One of authors (EAK) want to thank my collaborators Viktor
Sergeevich Fadin and Lev Nikolaevich Lipatov for fruitful
collaboration years ago. We are grateful to Azad Ahmadov for help.
One of us (EAK) is grateful to Heisenberg-Landau 2011 fund for
financial support. Besides EAK remind the 1978 year letter of LNL to
him "Greetings, Ed! About the problem of annihilation $e^+e^-$ to
quarks and gluons. The important region of integration in process
$e^+e^-\to q\bar{q}+ng$ depends on the kinematics of gluons emitted.
Every new gluon moves in the direction of quark (anti-quark) or in
the direction of parent quark, having the energy much less the one
of it's parent. Every gluon have it's genealogical tree...". In this
way the correlations in the polar angle can be investigated. The
authors express deep recognition to V.~A.~Nikitin for stimulating
discussions of these studies.

\end{document}